\documentclass[%
 reprint,
 amsmath,amssymb,
 aps,
]{revtex4-1}

\usepackage{graphicx}
\usepackage{float}
\usepackage{braket}
\usepackage{subcaption}
\usepackage{color, ulem}
\usepackage{gensymb}

\begin{document}

\title{Twist-angle sensitivity of electron correlations in moir\'e graphene bilayers}

\author{Zachary A. H. Goodwin}
\author{Fabiano Corsetti}
\author{Arash A. Mostofi}
\author{Johannes Lischner}
\affiliation{Departments of Materials and Physics and the Thomas Young Centre for Theory and Simulation of Materials, Imperial College London, South Kensington Campus, London SW7 2AZ, UK\\}

\date{\today}

\begin{abstract}
Motivated by the recent observation of correlated insulator states and unconventional superconductivity in twisted bilayer graphene, we study the dependence of electron correlations on the twist angle and reveal the existence of strong correlations over a narrow range of twist-angles near the magic angle. Specifically, we determine the on-site and extended Hubbard parameters of the low-energy Wannier states using an atomistic quantum-mechanical approach. The ratio of the on-site Hubbard parameter and the width of the flat bands, which is an indicator of the strength of electron correlations, depends sensitively on the screening by the semiconducting substrate and the metallic gates. Including the effect of long-ranged Coulomb interactions significantly reduces electron correlations and explains the experimentally observed sensitivity of strong correlation phenomena on twist angle.
\end{abstract}

\maketitle

\textit{Introduction}-The recent discovery of strong-correlation phenomena in magic-angle twisted bilayer graphene (tBLG), namely unconventional superconductivity in proximity to insulator states~\cite{NAT_I,NAT_S,TSTBLG,SMTBLG,SOM}, has generated tremendous interest~\cite{PSIF,IMACP,KVB,SCDID,EPRG,SCPKV,KL,CSD,SCHFC,WC, EPC,OMIB,NCIS,PMS,TCSSV,PCIS,US,EE,SMLWF,MMIT,MLWO}. The measured phase diagram of tBLG resembles that of cuprates~\cite{NAT_S,SMTBLG}, but the microscopic origin of the correlated states remains controversial~\cite{PSIF,IMACP,KVB,SCDID,EPRG,SCPKV,KL,CSD,SCHFC,WC, EPC,OMIB,NCIS,PMS,PCIS,TCSSV,US,EE}. tBLG offers unique advantages for studying strong electron correlations as it is highly tunable through the twist angle~\cite{GBWT,MBTBLG,LDE}, hydrostatic pressure~\cite{TSTBLG, PDTBLG}, doping, electric and magnetic fields and temperature~\cite{NAT_I,NAT_S,TSTBLG,SMTBLG,SOM}. Experimental measurements on tBLG, however, are highly sample dependent indicating a strong twist-angle sensitivity of strong correlation phenomena~\cite{NAT_I,NAT_S,TSTBLG,SOM}.

tBLG consists of two vertically-stacked graphene sheets that are rotated with respect to each other resulting in a moir\'e pattern that is generally incommensurate but, for certain angles, exhibits long-range periodicity associated with the moir\'e superlattice~\cite{GBWT,LDE,MBTBLG,NSCS,AC}. Theoretical studies show that at a ``magic" twist-angle of $\sim 1.1$\degree, around which the moir\'e unit cells associated with commensurate structures contain thousands of atoms, the width of the four bands near the Fermi level becomes very small~\cite{MBTBLG,LDE}, reflecting a reduction of the electronic kinetic energy. It is then expected that the ratio of the electron interaction energy to the electron kinetic energy increases, signalling the increasing dominance of electron-electron interactions and the emergence of strongly-correlated electronic behaviour~\cite{NAT_I,NAT_S,TSTBLG,SOM}. Indeed, correlated-insulator states and unconventional superconductivity are found when the system is doped by integer numbers of electrons or holes per moir\'e unit cell~\cite{NAT_I,NAT_S,TSTBLG,SMTBLG,SOM}. 

To help understand the microscopic origins of strong-correlation phenomena in tBLG, a wide range of theoretical approaches have been used. Atomistic tight-binding calculations~\cite{PDTBLG,LDE,NSCS,FBST} and continuum models~\cite{GBWT,MBTBLG,FTBM,OATBLG,CIC,BSS,OMACM,LDLE,CMLD,KDP,MLWO} have provided valuable insights into the band structure of tBLG, but do not capture the effect of electron correlations. The effect of electron-electron interactions have been studied using quantum Monte Carlo~\cite{PSIF,IMACP,KVB}, renormalization group~\cite{SCDID,EPRG,SCPKV,KL,CSD}, self-consistent Hartree-Fock~\cite{SCHFC}, and other theoretical and computational approaches~\cite{WC,NCIS,PCIS}.

The material-specific parameters that enter the interacting low-energy Hamiltonians of tBLG are often expressed in a Wannier function basis~\cite{OMIB,SCPKV}. Wannier functions (WFs) of tBLG have been constructed by Koshino \textit{et al.}~\cite{MLWO} using a continuum model and by Kang and Vafek~\cite{SMLWF} within atomistic tight-binding. These groups also used the WFs to calculate hopping parameters and Coulomb interaction matrix elements at a single twist angle near the magic angle~\cite{MLWO,SCPKV}. 

In this article, we investigate the dependence of electron correlations in tBLG on the twist angle. In particular, we carry out atomistic tight-binding calculations for a set of twist angles and construct WFs for each twist angle to determine the matrix elements of the screened Coulomb interaction between electrons in the flat bands. We demonstrate that both screening and the long-ranged interaction drastically reduce the range of twist-angles over which strong correlation phenomena may be expected. Specifically, the range is found to be only 0.1\degree~around the magic angle, in good agreement with experimental estimates~\cite{NAT_I,NAT_S,TSTBLG}.

\begin{figure}[h!]
\begin{center}
\includegraphics[width = 1\linewidth]{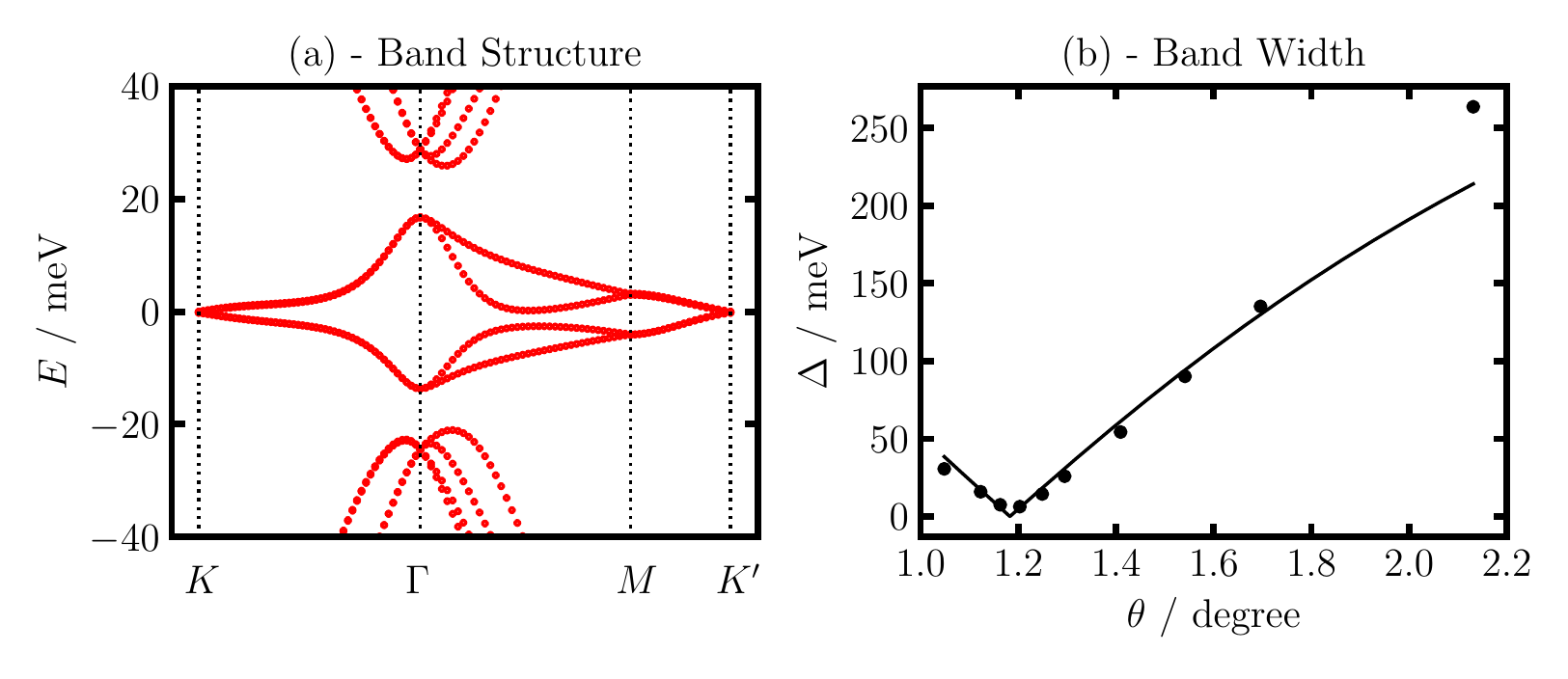}\\
\caption{(a) Atomistic tight-binding band structure for a twist angle of 1.05\degree\ (the Fermi level is at 0 meV). (b) Calculated band width as function of twist angle (dots) and analytical fit (solid black line).}
\label{FIG1}
\end{center}
\end{figure}

\textit{Methods}-To gain insights into the electronic structure of tBLG, we solve the atomistic tight-binding Hamiltonian given by
\begin{equation}
\mathcal{\hat{H}}_{0} = \sum_{i,j} \left\{ t(\textbf{r}_{i} - \textbf{r}_{j})\hat{c}^{\dagger}_{j}\hat{c}_{i} + \text{H.c.} \right\},
\end{equation}
where $\hat{c}^{\dagger}_{i}$ and $\hat{c}_{i}$ are, respectively, the creation and annihilation operators of electrons in $p_z$-orbitals of atom $i$, and $t(\textbf{r}_{i} - \textbf{r}_{j})$ is the hopping parameter between atoms $i$ and $j$ obtained using the Slater-Koster approach~\cite{LDE,NSCS,SK}. The effect of out-of-plane atomic corrugation~\cite{AC,LSDFT,PDTBLG,SETLA,STBBG} is included following Ref.~\citenum{MLWO}. See Supplementary Material (SM) for additional details.

Figure~\ref{FIG1}(a) shows the tight-binding band structure of tBLG at a twist angle of 1.05\degree. In good agreement with the literature~\cite{KDP,MLWO,PDTBLG,EDS,CTBS}, we find a set of four flat bands near the Fermi level. Fig.~\ref{FIG1}(b) shows the width $\Delta$ of the flat bands as function of the twist angle. The calculated band widths are accurately described by $\Delta = \delta \cdot (\theta^{2} - (\theta^{*})^{2})/(\theta^{2} + 2(\theta^{*})^{2})$ with a magic angle of $\theta^*=1.18$\degree and $\delta=0.5$ eV~\cite{MBTBLG}. Note that $\theta^*$ is slightly larger than that found in previous continuum model results~\cite{MBTBLG}.

As the flat bands are separated from all other bands by energy gaps in the magic-angle regime, maximally localized Wannier functions (MLWFs)~\cite{MAVAN,MLWF} can be constructed for these bands (without having to use a subspace selection procedure) according to
\begin{equation}
w_{n\textbf{R}}(\textbf{r}) = \dfrac{1}{\sqrt{N}}\sum_{m\textbf{k}}e^{-i\textbf{k}\cdot\textbf{R}}U_{mn\textbf{k}}\psi_{m\textbf{k}}(\textbf{r}),
\label{eq:wannier}
\end{equation}
where $w_{n\textbf{R}}$ is the WF and $\psi_{m\textbf{k}}$ denotes a Bloch eigenstate of the Hamiltonian with band index $m$ and crystal momentum $\textbf{k}$; $N=30\times 30$ is the number of $\textbf{k}$-points used to discretize the first Brillouin zone; \textbf{R} is a moir\'e lattice vector; $U_{mn\textbf{k}}$ is a unitary matrix that mixes the Bloch bands at each $\textbf{k}$ and represents the gauge freedom of the Bloch states. To obtain MLWFs, $U_{mn\textbf{k}}$ is chosen such that the total quadratic spread of the resulting WFs is minimised~\cite{MAVAN,MLWF}. 

\begin{figure}[h!]
\begin{center}
\includegraphics[width = 1\linewidth]{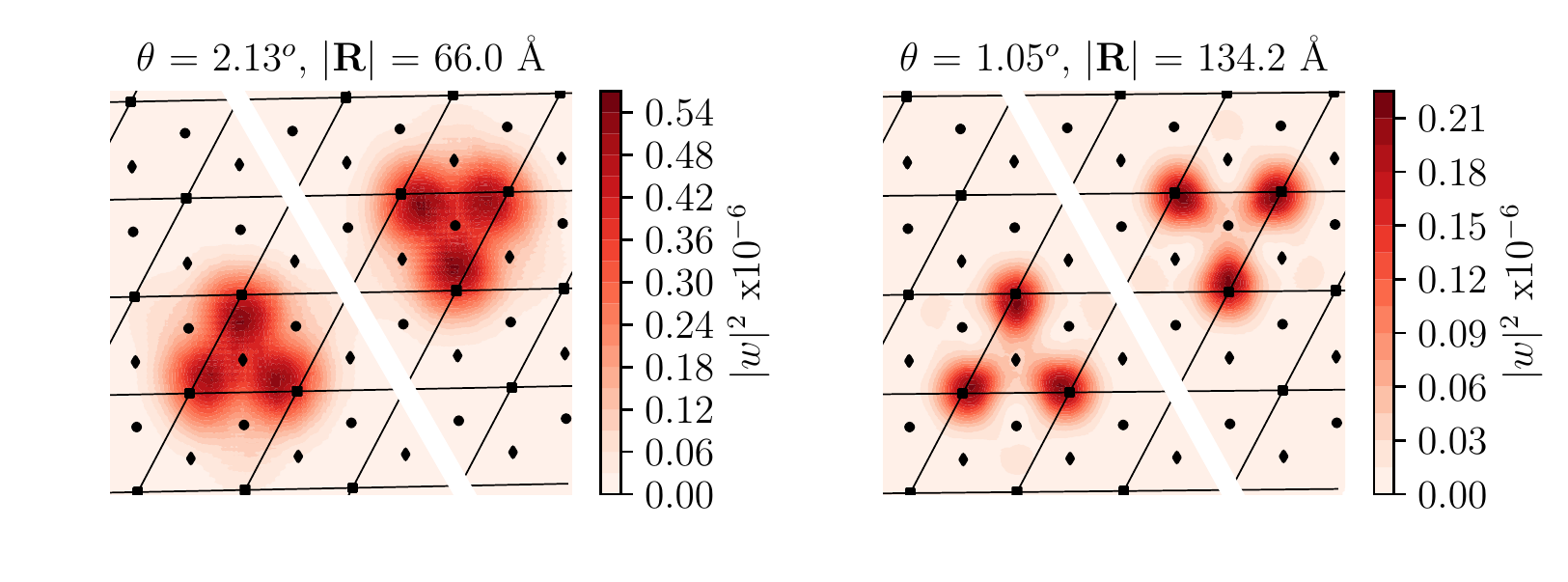}\\
\caption{Flat-band Wannier functions of tBLG with a twist angle of 2.13\degree\ (left) and 1.05\degree\ (right). Shown is the square modulus of the coefficients of the Wannier functions on each carbon atom. The squares, diamonds and circles denote the centers of the AA, AB and BA regions of tBLG, respectively.}
\label{FIG2}
\end{center}
\end{figure}

To obtain a Wannier-transformed Hamiltonian that reproduces the symmetries of the band structure of tBLG, the WFs must be centered at the AB or the BA positions of the moir\'e unit cell~\cite{MLWO,SMLWF,OMIB,MMIT} (shown in Fig.~\ref{FIG2}). We therefore use the approach of Ref.~\citenum{SLWF} and selectively localize two WFs and constrain the centres, one on each of these positions (see SM for more details).

To calculate MLWFs (see SM), it is expedient to choose an initial gauge by projecting the Bloch states onto some trial guess for the WFs~\cite{MAVAN,MLWF}. We tested two different starting guesses following suggestions from Ref.~\citenum{SMLWF} and Ref.~\citenum{MLWO}. Both initial guesses produce MLWFs with nearly identical shapes and the resulting Coulomb matrix elements differ by less than five percent (see SM for more details). In both cases, we obtain MLWFs using the Wannier90 code (version 3.0)~\cite{W90vT} with a custom interface to our in-house atomistic tight-binding code~\cite{FC}. Fig.~\ref{FIG2} shows the resulting MLWFs for two twist angles. In agreement with previous work~\cite{MLWO,SMLWF,OMIB}, we find the WFs exhibit three lobes that sit on the AA regions of the moir\'e unit cell.

In the Wannier basis, the interacting part of the Hamiltonian is given by 
\begin{equation}
\hat{\mathcal{H}}_{int} = \dfrac{1}{2}\sum_{\{n_{i}\textbf{R}_{i}\}}V_{\{n_{i}\textbf{R}_{i}\}}\hat{c}^{\dagger}_{n_{4}\textbf{R}_{4}}\hat{c}^{\dagger}_{n_{3}\textbf{R}_{3}}\hat{c}_{n_{2}\textbf{R}_{2}}\hat{c}_{n_{1}\textbf{R}_{1}},
\end{equation}
where $\hat{c}^{\dagger}_{n\textbf{R}}$ and $\hat{c}_{n\textbf{R}}$ are, respectively, the creation and annihilation operators of electrons in Wannier state $\ket{w_{n\textbf{R}}}$, and $V_{\{n_{i}\textbf{R}_{i}\}}$ denotes a matrix element of the screened Coulomb interaction, $W(\textbf{r}-\textbf{r}^{\prime})$. The largest matrix elements are usually obtained when $\textbf{R}_{4} = \textbf{R}_{1}$, $\textbf{R}_{3} = \textbf{R}_{2}$, $n_4=n_1$ and $n_3=n_2$. For this case, the Coulomb matrix element is given by 
\begin{equation}
V_{ij} = \iint d\textbf{r}d\textbf{r}^{\prime}|w_{i}(\textbf{r})|^{2}W(\textbf{r} - \textbf{r}^{\prime})|w_{j}(\textbf{r}^{\prime})|^{2}.
\label{GHI}
\end{equation}

We evaluate Eq.~\eqref{GHI} for two models of the screened interaction. In the first case a Coulomb potential is used, $W(r) = e^{2}/4\pi\epsilon_{r} \epsilon_{0}r$. The dielectric constant $\epsilon_r$ has contributions from the substrate (typically hBN~\cite{NAT_I,NAT_S,TSTBLG}) and high-energy bands of tBLG. Values between 6 and 10 have been used in the literature~\cite{SCPKV,WC}; here, we use $\epsilon_r=8$.

In the second case, we include the effect of metallic gates on both sides of the tBLG (but separated from it by the hBN substrate). The resulting screened interaction is given by~\cite{MGS}
\begin{equation}
W^{g}(\textbf{r} -\textbf{r}^{\prime}) = \dfrac{e^{2}}{4\pi\epsilon_{r}\epsilon_{0}}\sum_{n=-\infty}^{+\infty}\dfrac{(-1)^{n}}{\sqrt{|\textbf{r} -\textbf{r}^{\prime}|^{2} + (\xi n)^{2}}},
\label{eq:Wgate}
\end{equation}
where $\xi=10$~nm is half the distance between the two metallic gates~\cite{MGS,SCPKV}. For $|\textbf{r} -\textbf{r}^{\prime}| \ll \xi$, $W^{g}$ is proportional to the bare Coulomb interaction ($n$=0 term). In the opposite limit, the interaction simplifies to $W^{g}(r) = \sqrt{2}e^{2}e^{-\pi r/\xi}/(2\pi\epsilon_{r}\epsilon_{0}\sqrt{r\xi}$)~\cite{MGS}. See SM for more details. 

\begin{figure*}[t!]
\begin{center}
\includegraphics[width = 1\linewidth]{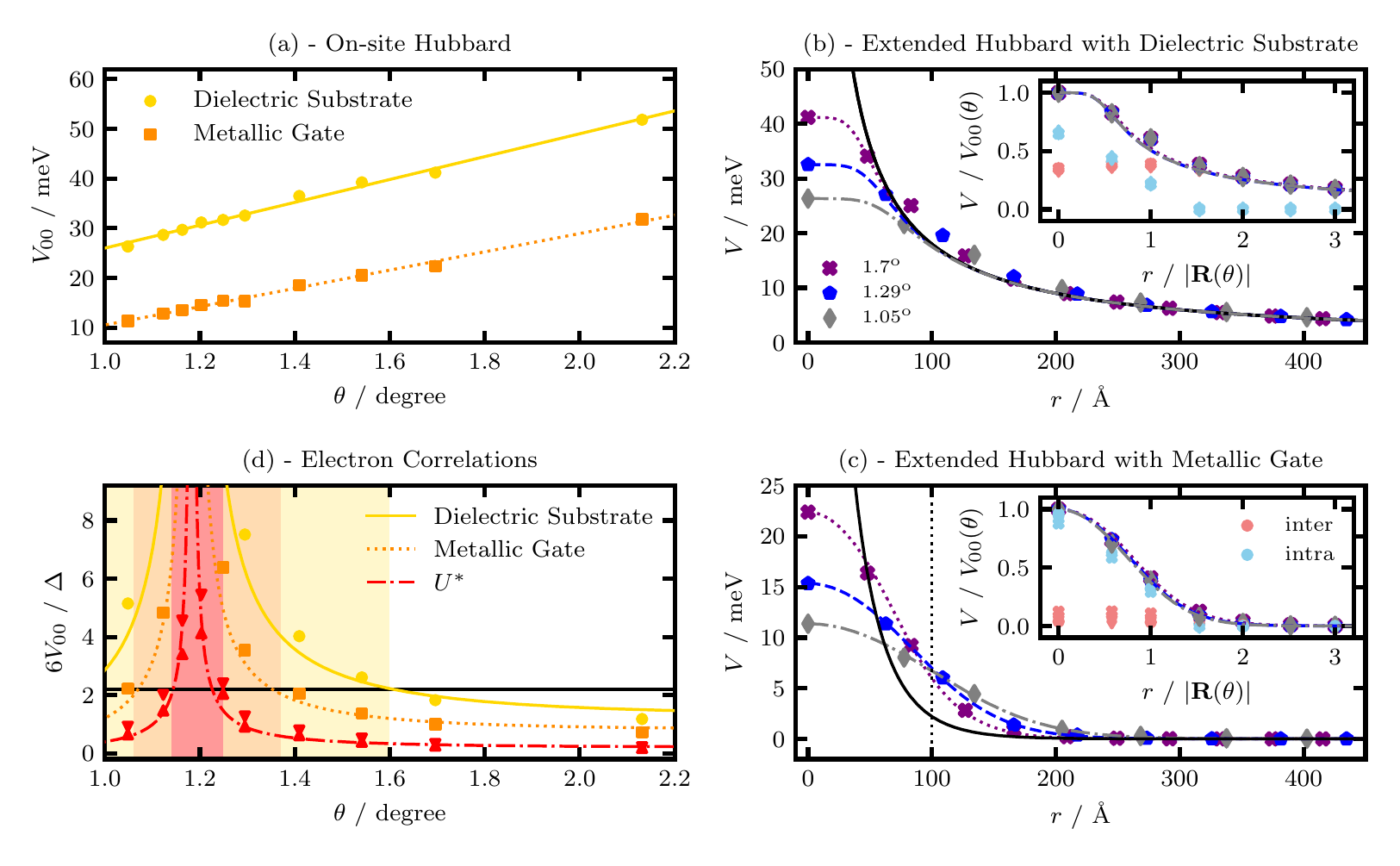}\\
\caption{(a) On-site Hubbard parameter in tBLG encapsulated by hBN as function of twist angle for different models of the screened interaction. (b) Extended Hubbard parameters for tBLG on hBN as a function of distance between Wannier functions for three angles (1.70\degree, 1.29\degree\ and 1.05\degree); dotted lines represent Eq.~\eqref{OHNO_C} and solid line dielectric screened Coulomb potential. Inset: rescaled extended Hubbard parameters and contributions from intra-lobe (cyan symbols) and inter-lobe (red symbols) contributions. (c) Extended Hubbard parameters for tBLG on hBN with metallic gates as a function of the distance between Wannier orbitals; dotted lines represent Eq.~\ref{OHNO_M} and solid line large separation limit of Eq.~\eqref{eq:Wgate}. Inset: rescaled extended Hubbard parameters and decomposition into intra- and inter-lobe contributions. (d) Ratio of the on-site Hubbard interaction to band width in tBLG as function of twist angle. Circles denote results for tBLG on hBN; squares denote results for tBLG on hBN with metallic gates; and triangles denote results for tBLG on hBN with (upwards facing triangle) and without (downwards facing triangle) metallic gates when extended Hubbard parameters are taken into account.}
\label{FIG3}
\end{center}
\end{figure*}

\textit{Results and Discussion}-The circle data points in Fig.~\ref{FIG3}(a) show the on-site Hubbard parameter, $V_{00}$, of tBLG without metallic gates as function of twist angle. In this range of twist angles, the on-site Hubbard parameters have values of approximately $25 \sim 50$~meV, two orders of magnitude smaller than in graphene~\cite{SECI}. Moreover, we find that $V_{00}$ depends approximately linearly on the twist angle, i.e., $V_{00} = (m_{00}\theta+c_{00})/\epsilon_{r}$ with $m_{00} = 200$ meV/degree and $c_{00}=24$~meV. This dependence can be understood from the following scaling argument. If the decay length of the WF is proportional to the size of the moir\'e unit cell (and the WFs have no other twist-angle dependence), transforming the integrals in Eq.~\eqref{GHI} to dimensionless coordinates immediately shows that $V_{00}$ scales as the inverse size of the moir\'e unit cell length which is inversely proportional to $\theta$ in the limit of small twist angles.

Including the screening from the metallic gates reduces the on-site Hubbard parameter by roughly a factor of two, see squares in Fig.~\ref{FIG3}(a). Again, we find a linear dependence of $V^g_{00}$ on the twist angle with $m^g_{00}=148$~meV/degree and $c^g_{00}=-63$~meV. This finding is surprising as the scaling argument applied to the screened interaction of Eq.~\eqref{eq:Wgate} suggests that the resulting dimensionless integrals should be strong functions of $\theta$. We expect that this non-linear behavior would be seen over a larger range of twist angles than that studied here.

Figure~\ref{FIG3}(b) shows the extended Hubbard parameters of tBLG without metallic gates as function of the separation between WF centres for three twist angles. The extended Hubbard parameters decay slowly as function of distance as a consequence of the long-ranged Coulomb interaction and converge to the screened interaction evaluated at the Wannier centres for distances larger than four moir\'e unit cells [black solid line in Fig.~\ref{FIG3}(b)].

We fit our results for the extended Hubbard parameters (including the on-site term) to a modified Ohno potential~\cite{Ohno} 
\begin{equation}
V(r,\theta) = \dfrac{V_{00}(\theta)}{\sqrt[5]{1 + \big(V_{00}(\theta)/W(r)\big)^{5}}}.
\label{OHNO_C}
\end{equation}
Fig.~\ref{FIG3}(b) shows, as seen with the dotted lines, that this expression accurately describes the calculated extended Hubbard parameters for \textit{all twist angles} with only two parameters, $m_{00}$ and $c_{00}$ (see SM for more details). The inset of Fig.~\ref{FIG3}(b) shows that the extended Hubbard parameters collapse onto a universal \textit{twist-angle independent} curve when the WF separation is divided by the moir\'e unit cell length $|\textbf{R}(\theta)|$.

The inset of Fig.~\ref{FIG3}(b) also shows the contributions to the extended Hubbard parameters from intra- and interlobe interactions of the WFs~\cite{MLWO}. The intra-lobe contributions decay to zero after second nearest neighbours, while the inter-lobe contributions initially increase (as a consequence of having more non-overlapping lobe pairs) and then decay slowly.

Figure~\ref{FIG3}(c) shows that when screening from the metallic gates is taken into account, the extended Hubbard parameters decay to zero on a length scale of the order of the tBLG--gate distance, $\xi = 10$nm (dotted vertical line). These extended Hubbard parameters can be accurately described by the modified Ohno potential of Eq.~\eqref{OHNO_C} multiplied by a Gaussian
\begin{equation}
V^{g}(r,\theta) = \dfrac{V^{g}_{00}(\theta)e^{-(r/\alpha|\textbf{R}(\theta)|)^{2}}}{\sqrt[5]{1 + \big(V^{g}_{00}(\theta)/W^g(r) \big)^{5}}}.
\label{OHNO_M}
\end{equation}
We find $\alpha \sim 1.1$ provides a good description of the data in the range of twist angles studied. 

Again, the extended Hubbard parameters collapse onto a universal curve upon rescaling the distances, as shown in the inset of Fig.~\ref{FIG3}(c). The inset also shows that the extended Hubbard parameters are dominated by intra-lobe contributions as the finite range of $W^g$ reduces the contribution from inter-lobe terms. This observation also explains the reduction of the on-site Hubbard parameter by a factor of two in the presence of metallic gates [Fig.~\ref{FIG3}(a)]: without metallic gates, approximately half of $V_{00}$ is contributed by inter-lobe interactions which are screened out by the gates.

Figure~\ref{FIG3}(d) shows the ratio of the on-site Hubbard parameter $V_{00}$ to the band width $\Delta$ as function of twist angle for different screened interactions. Note that we have multiplied $V_{00}/\Delta$ by a factor of six to approximate $V_{00}/t$ which is typically used to characterize the strength of electronic correlations ($\Delta=6t$ for graphene with nearest-neighbour hopping only \cite{EPG}). As expected, $V_{00}/t$ becomes large near the magic angle. The largest values of $V_{00}/t$ are obtained for the screened interaction without metallic gates. Taking the screening from the metallic gates into account reduces $V_{00}/t$ by approximately a factor of two.

Our results thus demonstrate that electron correlations in tBLG can be continuously tuned as function of twist angle from a weakly correlated to a strongly correlated regime in the vicinity of the magic angle. Calculating the phase diagram of such a system is extremely challenging as most theoretical approaches are tailored to one of the two limiting cases and are correspondingly classified as weak-coupling or strong-coupling techniques. Quite generally, it is expected that tBLG undergoes a metal-to-insulator transition as the strength of electron correlations increases, but the detailed microscopic nature of the insulating phase remains controversial. 

Mean-field theory and strong coupling techniques predict that the gapped phase in undoped tBLG is an antiferromagnetic insulator~\cite{ECM,KVB,IMACP,TJMODEL}. However, the exact value of the critical $V_{00}/t$ where the transition occurs has not been established. For (untwisted) Bernal stacked bilayer graphene, accurate Quantum Monte Carlo calculations yield a critical value of $6V_{00}/\Delta = 2.2$~\cite{PSIF,ABS} [black horizontal line in Fig.~\ref{FIG3}(d)]. Without metallic gates, we find that the electronic correlations in tBLG exceed this critical value in a relatively large twist-angle range (from angles smaller than $\theta=1.0\degree$ up to $\theta=1.6\degree$). With metallic gates, the critical twist-angle range is reduced by over a factor of two (from $\theta=1.06\degree$ to $\theta=1.37\degree$).

In materials with significant, long-ranged Coulomb interactions, a different measure of strong correlations is appropriate. In particular, in such systems the energy gained by moving one electron from a doubly-occupied orbital to a neighboring orbital is not $V_{00}$, but $U^*=V_{00}-V_{01}$~\cite{OHP}. As $U^*$ is about three (five) times smaller than $V_{00}$ for the case of screening with (without) a metallic gate, long-range interactions drastically reduce the window of strongly correlated twist angles [see red curve in Fig.~\ref{FIG3}(d) which is calculated from both interaction potentials studied here and found to be essentially independent of the type of interaction]. In particular, we find that the width of the critical twist-angle window is only $0.1\degree$ which is in good agreement with recent experimental findings and explains the observed sensitivity of experimental measurements to sample preparation~\cite{NAT_I,NAT_S,TSTBLG,SOM}. 

While gapped states in tBLG have been observed at charge neutrality~\cite{SOM}, there is also significant interest in correlated insulator states of electron- or hole-doped systems~\cite{NAT_I,NAT_S,TSTBLG,SOM}. Away from charge neutrality, weak coupling approaches predict a transition from a metallic to a gapped antiferromagnetic phase at specific values of the Fermi level when the Fermi surface exhibits nesting with a critical value of $U^*/t \sim 2$~\cite{SCDID}. This suggests that the width of the strongly correlated twist-angle window does not depend sensitively on doping and again highlights the importance of the extended Hubbard parameters. In contrast, strong coupling calculations of doped tBLG predict that gapped \textit{ferromagnetic} spin- or valley-polarized ground states occur whenever the number of additional carriers per moir\'e unit cell is integer~\cite{SCHFC}. This suggests the intriguing possibility that \textit{multiple} phase transitions occur within the narrow, strongly correlated, twist-angle window.

Superconductivity in tBLG occurs at low temperatures in the vicinity of the correlated insulator phases~\cite{NAT_S,TSTBLG,SOM}. While some works have suggested phonons as being responsible for the pairing mechanism~\cite{PMS,EPC}, similarities to the cuprate phase diagram indicate that non-phononic mechanisms could be relevant in tBLG~\cite{SCDID,KL,TJMODEL,CSD}. For example, superconductivity emerges in weak-coupling approaches from the exchange of damped spin waves~\cite{SCDID,KL,CSD}. Gonazalez and Stauber~\cite{KL} have shown that very small values of $U^*/t$ are sufficient to trigger superconductivity when the Fermi level lies near the van Hove singularity. This suggests that superconductivity should be observable in a larger twist-angle range than the correlated insulator phases.  

\textit{Summary}-We studied the twist-angle dependence of electron correlations in tBLG. For this, we calculated on-site and extended Hubbard parameters for a range of twist angles and demonstrated that the on-site Hubbard parameters depend linearly on twist angle for both dielectric substrate and metallic gate screened interaction potentials. The extended Hubbard parameters decay slowly as function of the Wannier function separation and are reproduced accurately \textit{for all twist angles} with an Ohno-like potential. By calculating the ratio of the interaction energy and the kinetic energy of electrons in tBLG, we predict the twist-angle windows where strong correlation phenomena occur. When the reduction of electron correlations arising from both screening and the long range of the electron interaction are taken into account, we find a critical twist-angle window of only $0.1\degree$ which explains the experimentally observed twist-angle sensitivity of strong correlation phenomena in tBLG.

\textit{Acknowledgements}-We thank V. Vitale, D. Kennes, C. Karrasch and A. Khedri for helpful discussions. This work was supported through a studentship in the Centre for Doctoral Training on Theory and Simulation of Materials at Imperial College London funded by the EPSRC (EP/L015579/1). We acknowledge funding from EPSRC grant EP/S025324/1 and the Thomas Young Centre under grant number TYC-101.

\bibliographystyle{apsrev4-1}
\bibliography{ADI}

\appendix

\renewcommand{\theequation}{S\arabic{equation}}
\renewcommand{\thefigure}{S\arabic{figure}}
\setcounter{figure}{0} 

\onecolumngrid

\section{Methods}

\subsection{Atomic structure}

\begin{figure*}[t!]
\begin{center}
\includegraphics[width = 1\linewidth]{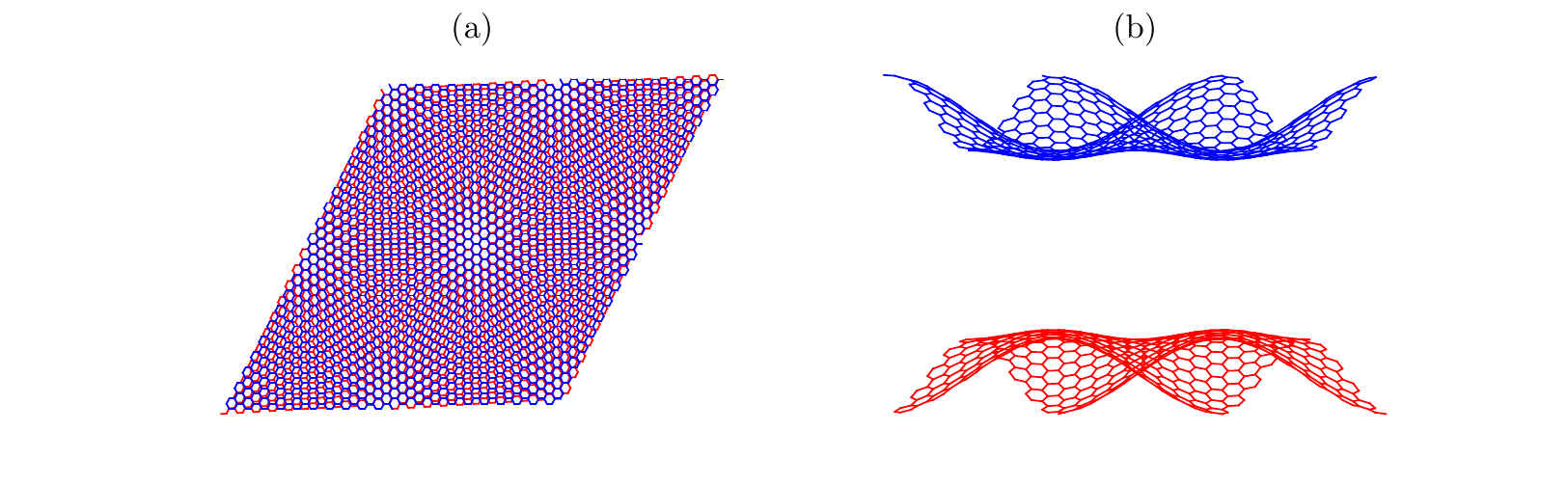}\\
\caption{(a) Moir\'e lattice of twisted bilayer graphene for a twist angle of 3.15$^{\text{o}}$. (b) Side view of twisted bilayer graphene. Note that the atomic structure was calculated with $d_{AB} = 1$ $\textrm{\AA}$ and $d_{AB} = 2$ $\textrm{\AA}$ with the z-axis expanded for viewing purposes.}
\label{SF_FIG}
\end{center}
\end{figure*}

The structure of twisted bilayer graphene (tBLG) is generated from AA stacked bilayer graphene by rotating the top graphene sheet around an axis perpendicular to the bilayer that intersects one carbon atom in each layer, producing a structure with D$_{3}$ symmetry~\cite{LDE}. To obtain a commensurate structure, see Fig.~\ref{SF_FIG}(a) for example, an atom of the rotated top layer must reside above an atom of the (unrotated) bottom layer. The corresponding lattice vectors of the moir\'e unit cell are given by $\textbf{R}_{1} = n\textbf{a}_{1} + m \textbf{a}_{2}$ and $\textbf{R}_{2} = -m\textbf{a}_{1} + (n + m) \textbf{a}_{2}$, where $n$ and $m$ are integers and $\textbf{a}_{1} = (\sqrt{3}/2, -1/2)a_{0}$ and $\textbf{a}_{2} = (\sqrt{3}/2, 1/2)a_{0}$ denote the lattice vectors of graphene with $a_{0} = 2.46$ $\textrm{\AA}$ being graphene's lattice constant~\cite{LDE,NSCS}. The twist angle $\theta$ can be obtained from $n$ and $m$ via

\begin{equation}
\cos\theta = \dfrac{n^{2} + 4nm + m^{2}}{2(n^{2} + nm + m^{2})}.
\end{equation}

At small twist angles ($\theta < 10^{\text{o}}$), significant lattice relaxations occur in tBLG~\cite{AC,LSDFT,PDTBLG,SETLA,STBBG}. There are both in-plane relaxations, resulting from the growth of the lower-energy AB regions and corresponding shrinkage of AA regions, and large out-of-plane corrugations arising from the different interlayer separations of AB and AA stacked bilayer graphene~\cite{SETLA,STBBG}, see Fig.~\ref{SF_FIG}(b). Here, we only take out-of-plane relaxations into account as they have a larger magnitude than in-plane distortions. Specifically, we employ the expression proposed in Ref.~\citenum{MLWO} for the vertical displacement of carbon atoms at position $\textbf{r}$ given by 
\begin{equation}
z(\textbf{r}) = d_{0} + 2d_{1}\sum_{i}\cos(\textbf{b}_{i}\cdot\textbf{r}).
\end{equation}
Here, the summation runs over the primitive reciprocal lattice vectors of tBLG, $\textbf{b}_{1/2}$, and the sum of these vectors; and $d_{0} = (d_{AA} + 2d_{AB})/3$ and $d_{1} = (d_{AA} - d_{AB})/9$ with $d_{AB} = 3.35$ $\textrm{\AA}$ and $d_{AA} = 3.60$ $\textrm{\AA}$ being the interlayer separations of AB and AA stacked bilayer graphene, respectively~\cite{MLWO}. 

\subsection{Slater-Koster Rules}

To calculate the hopping parameters, we employ the Slater-Koster approach~\cite{SK,LDE,FC}
\begin{equation}
t(\textbf{r}) = V_{pp\sigma}(\textbf{r})\bigg(\dfrac{\textbf{r}\cdot\textbf{e}_{z}}{|\textbf{r}|}\bigg)^{2} + V_{pp\pi}(\textbf{r})\bigg(1 - \bigg[\dfrac{\textbf{r}\cdot\textbf{e}_{z}}{|\textbf{r}|}\bigg]^{2}\bigg);
\end{equation}
where $V_{pp\sigma}(\textbf{r}) = V_{pp\sigma}^{0}\exp\{q_{\sigma}(1 - |\textbf{r}|/d_{AB})\}$ and $V_{pp\pi}(\textbf{r}) = V_{pp\pi}^{0}\exp\{q_{\pi}(1 - |\textbf{r}|/a)\}$ with $V_{pp\sigma}^{0} = 0.48$ eV and $V_{pp\pi}^{0} = -2.7$ eV \cite{FC,SK,EPG}. Note that $a = 1.42$ $\textrm{\AA}$ is the carbon-carbon bond length in graphene and $q_{\sigma} = 7.43$ and $q_{\pi} = 3.14$~\cite{LDE,NSCS}. 

\subsection{Bloch States}

The Bloch eigenstates of the tight-binding Hamiltonian are given by 
\begin{equation}
 \psi_{n\textbf{k}}(\textbf{r}) = \frac{1}{\sqrt{N}}\sum_{j\textbf{R}}c_{jn\textbf{k}}e^{i\textbf{k}\cdot \textbf{R}}\phi_{z}(\textbf{r} - \textbf{t$_{j}$} - \textbf{R}),
\end{equation}
where $\phi_z$ denotes the wavefunction of the p$_z$-orbital, $\textbf{t$_{j}$}$ is the position of carbon atom $j$ in the unit cell, $N$ denotes the number of moir\'e unit cells in the crystal and $c_{jn\textbf{k}}$ are coefficients obtained from the diagonalization of the Hamiltonian. 

\subsection{Wannier Functions}

As mentioned in the main text, the four flat bands near the Fermi energy are separated form all other bands by energy gaps in the magic-angle regime. Hence, these bands form a manifold that can be wannierized without a disentanglement procedure~\cite{MLWF}. 

To constrain the Wannier function centres, the selective localization method was employed to calculate two Wannier functions: one centered on the AB position and the other on the BA position of the moir\'e unit cell~\cite{MLWO,SMLWF,OMIB,MMIT}. We utilise the approach of Wang \textit{et al.}~\cite{SLWF} and constrain the centres of two Wannier functions to lie at the AB and BA positions. In this approach, one minimizes the cost function
\begin{equation}
\Omega = \sum_{n=1}^{J^{\prime}}\Big[\braket{r^{2}}_{n} - \bar{\textbf{r}}_{n}^{2} + \lambda(\bar{\textbf{r}}_{n} - \textbf{r}_{0n})^{2}\Big],
\end{equation}
where the first two terms describe the quadratic spread of the Wannier functions (with $\braket{r^{2}}_{n} = \braket{w_{n\textbf{R}}|r^{2}|w_{n\textbf{R}}}$ and $\bar{\textbf{r}}_{n} = \braket{w_{n\textbf{R}}|\textbf{r}|w_{n\textbf{R}}}$~\cite{MAVAN,MLWF}) and the third term enforces the additional constraint that the centre of the $n$-th Wannier function should be located at position $\textbf{r}_{0n}$~\cite{SLWF}. Also, $\lambda=200$ denotes the cost parameter and we use $J'=2$.

\begin{figure*}[t!]
\begin{center}
\includegraphics[width = 0.7\linewidth]{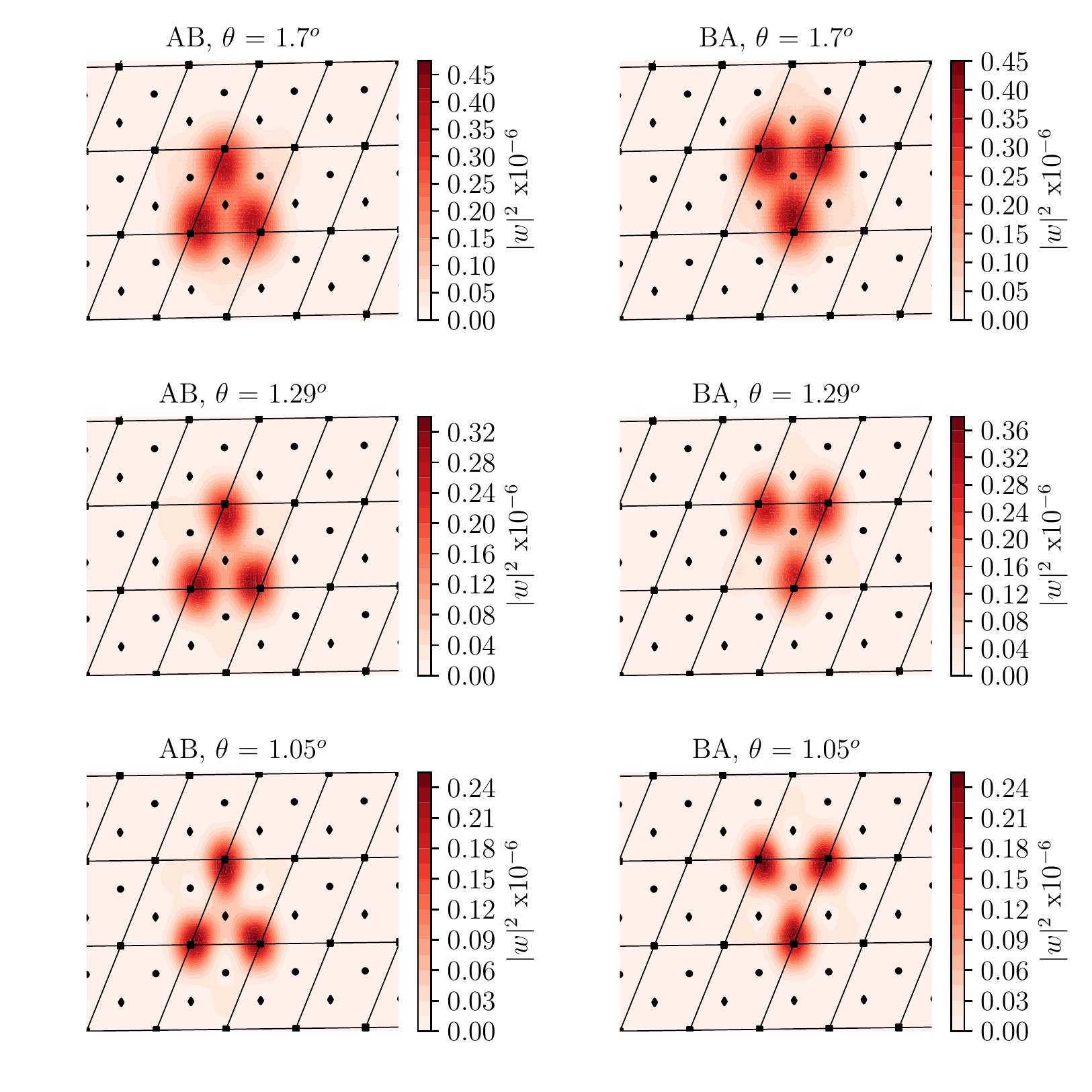}\\
\caption{Calculated Wannier functions centred on AB and BA positions for the three twist angles that were studied for the extended Hubbard parameters.}
\label{WFFIG}
\end{center}
\end{figure*}

To calculate maximally localized Wannier functions, a starting guess is required~\cite{MAVAN,MLWF}. We constructed two different starting guesses following suggestions from Ref.~\citenum{SMLWF} and Ref.~\citenum{MLWO}, and studied the dependence of the Wannier functions on the initial guess. For the first guess~\cite{SMLWF}, a linear combination of Bloch states at the $\Gamma$-point is constructed and then multiplied by a Gaussian envelope function with an appropriately chosen decay length. For the second guess~\cite{MLWO}, the gauge of Bloch states with a given band index was fixed by imposing that the wavefunctions are positive and real at either the AB or the BA positions. Then, the resulting Bloch states were inserted into Eq. (2) of the main text and transformed with $U^\textbf{k}_{nm}=\delta_{nm}$. For both starting points, we determine maximally localized Wannier functions using a $30\times 30\times 1$ k-point grid as implemented in the Wannier90 code (version 3.0)~\cite{W90vT}. We find that the final Wannier functions from the two initial guesses are qualitatively very similar to each other. In Fig.~\ref{WFFIG}, Wannier functions for the initial guess of Ref.~\citenum{SMLWF} can be seen for three twist angles.

\subsubsection{Input Calculation Details}

We are required to calculate $M_{mn}^{\textbf{k,q}} = \braket{u_{m\textbf{k}}|u_{n\textbf{k}+\textbf{q}}}$ for the Wannier90 code~\cite{W90vT}, where $u_{n\textbf{k}}$ is the unit cell periodic part of the Bloch state, as seen by $\psi_{n\textbf{k}} = e^{i\textbf{k}\cdot\textbf{r}}u_{n\textbf{k}}(\textbf{r})$. Inserting these unit cell periodic functions and shifting coordinate systems with the transformation $\textbf{r}^{\prime} =\textbf{r} - \textbf{R}$, evaluating a sum and then assuming contributions only come from the overlap of the same orbital, we arrive at

\begin{equation}
\braket{u_{m\textbf{k}}|u_{n\textbf{k}+\textbf{q}}} = \sum_{j}c^{*}_{m\textbf{k}i}c_{n\textbf{k}+\textbf{q}i}e^{i\textbf{q}\textbf{t$_{j}$}}\int d\textbf{r$^{\prime\prime}$}\phi^{*}_{z}(\textbf{r$^{\prime\prime}$})e^{i\textbf{q}\textbf{r$^{\prime\prime}$}}\phi_{z}(\textbf{r$^{\prime\prime}$}),
\end{equation}

\noindent where $\textbf{r$^{\prime\prime}$} = \textbf{r$^{\prime}$} - \textbf{t$_{j}$}$. 

Here $\phi_{z}$ is the pseudo-hydrogenic $p_{z}$ orbital of carbon atoms. The integral

\begin{equation}
I(\textbf{q}) = \int d\textbf{r}\phi^{*}_{z}(\textbf{r})e^{i\textbf{q}\textbf{r}}\phi_{z}(\textbf{r}),
\end{equation}

\noindent can be solved exactly, yielding $I(\textbf{q}) = [1 + (|\textbf{q}|a_{0}/Z)^{2}]^{-3}$, where $a_{0}$ is the Bohr radius and $Z$ is the effective charge of the carbon atom, taken to be 3.18~\cite{Shung}. 

The initial guess, $g_{n}$, is utilised to calculate $A_{mn}^{\textbf{k}} = \braket{\psi_{m\textbf{k}}|g_{n}}$ for the Wannier90 code~\cite{W90vT}. In one of the guesses we fix the gauge of the Bloch state at each $\textbf{k}$-point, $\tilde{\psi_{n\textbf{k}}}$, and separately Fourier transform each state to yield $g_{n}$. With this guess, we simply have $A_{mn}^{\textbf{k}} = \delta_{mn}$. 

The other initial guess can be expressed in the form

\begin{equation}
\ket{g_{n}} = \sum_{n^{\prime}}\psi^{v}_{n^{\prime}\Gamma}(\textbf{r})f(\textbf{r} - \textbf{r}_{0}),
\end{equation}

\noindent where $f(\textbf{r} - \textbf{r}_{0})$ is a Gaussian function centred at $\textbf{r}_{0}$ and $v$ denotes a sub-lattice and layer of tBLG. Inserting this guess and the Bloch states in a local basis set, we have

\begin{equation}
A_{mn}^{\textbf{k}} = \dfrac{1}{N}\sum_{n^{\prime}}\sum_{\textbf{R}\textbf{R}^{\prime}}\sum_{jv_{i}}c^{*}_{m\textbf{k}j}c_{n^{\prime}\Gamma v_{i}}e^{-i\textbf{k}\cdot\textbf{R}}\int d\textbf{r}\phi^{*}(\textbf{r} - \textbf{t}_{j} - \textbf{R})f(\textbf{r} - \textbf{r}_{0})\phi(\textbf{r} - \textbf{t}_{v_{i}} - \textbf{R}^{\prime}).
\end{equation}

\noindent Note that $v_{i}$ only runs over the atoms located on the layer and sublattice of $v$. Let's assume that only non-vanishing contributions come from the same $p_{z}$ orbital, and that the Gaussian is a slowly varying function, such that it can be taken outside of the integral. After evaluating these operations, we arrive at

\begin{equation}
A_{mn}^{\textbf{k}} = \dfrac{1}{N}\sum_{n^{\prime}}\sum_{\textbf{R}}\sum_{v_{i}}c^{*}_{m\textbf{k}v_{i}}c_{n^{\prime}\Gamma v_{i}}e^{-i\textbf{k}\cdot\textbf{R}}f(\textbf{t}_{v_{i}} + \textbf{R} - \textbf{r}_{0}).
\end{equation}

\noindent This summation over $\textbf{R}$ is performed over the entire crystal. 

\subsection{Coulomb Matrix Elements}

To evaluate Eq. (4) of the main text, the Wannier functions are expressed as a linear combination of $p_{z}$-orbitals according to

\begin{equation}
w_{n\textbf{R}}(\textbf{r}) = \sum_{j\textbf{R}^{\prime}}c_{n\textbf{R}\textbf{R}^{\prime}j}\phi_{z}(\textbf{r} - \textbf{t$_{j}$} - \textbf{R}^{\prime})
\label{WFLBS}
\end{equation}

\noindent with

\begin{equation}
c_{n\textbf{R}\textbf{R}^{\prime}j} = \dfrac{1}{N}\sum_{m\textbf{k}}U_{nm}^{(\textbf{k)}}e^{i\textbf{k}(\textbf{R}^{\prime} - \textbf{R})}c_{m\textbf{k}j}.
\end{equation}

Inserting Eq.~\eqref{WFLBS} into Eq. (4) of the main text yields 

\begin{equation}
V_{n_{1}\textbf{R}_{1}n_{2}\textbf{R}_{2}} = \sum_{\textbf{R}^{\prime}\textbf{R}^{\prime\prime}}\sum_{lj} |c_{n_{1}\textbf{R}_{1}l\textbf{R}^{\prime}}|^{2}|c_{n_{2}\textbf{R}_{2}j\textbf{R}^{\prime\prime}}|^{2}v_{l\textbf{R}^{\prime}j\textbf{R}^{\prime\prime}},
\label{FE}
\end{equation}

\noindent where $v_{l\textbf{R}^{\prime}j\textbf{R}^{\prime\prime}}$ denotes the Coulomb matrix elements between pairs of $p_{z}$-orbitals at positions $\textbf{t}_l+\textbf{R}^{\prime}$ and $\textbf{t}_j+\textbf{R}^{\prime\prime}$, respectively. For in-plane separations larger the than nearest neighbor distance, we assume $v_{l\textbf{R}^{\prime}j\textbf{R}^{\prime\prime}}=W(\textbf{t}_l+\textbf{R}^{\prime}-[\textbf{t}_j+\textbf{R}^{\prime\prime}])$. For the on-site ($v_{00} = 17$ eV) and nearest neighbour ($v_{01} = 8.5$ eV) terms, we used values obtained from \textit{ab initio} DFT calculations~\cite{SECI}. Eq.~\eqref{FE} is evaluated by explicitly carrying out the summations in real space using a $5\times 5$ supercell which yields highly converged results. While evaluating $v_{l\textbf{R}^{\prime}j\textbf{R}^{\prime\prime}}$ is straightforward for the Coulomb interaction screened by a semiconducting substrate, the case of a metallic gate is more difficult. Here, we evaluate Eq. (5) of the Main text by summing contributions up to $n=8$ which was found to reasonably reproduce the fully converged potential well for distances smaller than 40~$\textrm{\AA}$. For larger distances, we employ the analytical long-distance limit of $W^g$, see discussion following Eq. (5) of the main text. These approximations result in errors of less than five percent in the Hubbard parameters.

\clearpage

\section{On-site Hubbard Parameters}

The following labels are used throughout this section to refer to different initial guesses and centres of the Wannier functions.
\begin{itemize}
    \item (1) - Initial guess from Ref. \citenum{MLWO} centred on AB position
    \item (2) - Initial guess from Ref. \citenum{MLWO} centred on BA position 
    \item (3) - Initial guess from Ref. \citenum{SMLWF} centred on AB position
    \item (4) - Initial guess from Ref. \citenum{SMLWF} centred on BA position
\end{itemize}

\begin{figure*}[t!]
\begin{center}
\includegraphics[width = 1\linewidth]{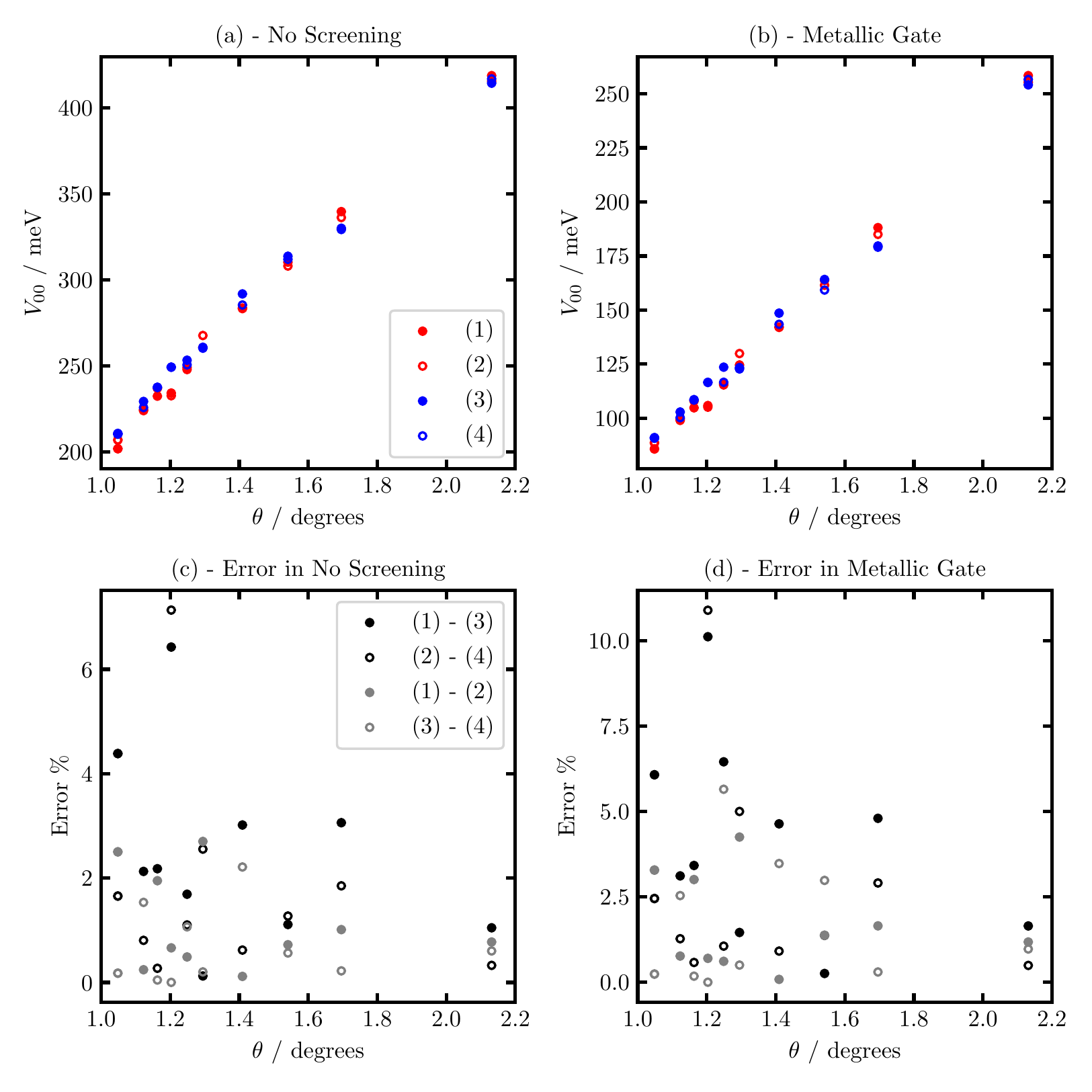}\\
\caption{For (a) and (b), the different symbols and colours represent different initial guesses as inputs for the selective localisation method of wannier90 (V3.0)~\cite{W90vT}. For details of these see the start of this section. For (c) and (d) there are two numbers for each symbol. These refer to the percentage difference between these two initial guesses from (a) and (b), respectively. (a) - On-site Hubbard parameter as a function of twist angle calculated with Coulomb interaction and $\epsilon_{r} = 1$. (b) - On-site Hubbard parameter as a function of twist angle calculated in the presence of a metallic gate and $\epsilon_{r} = 1$. (c) - Percentage errors between on-site Hubbard parameters at each twist angle for (a). (d) - Percentage errors between on-site Hubbard parameters at each twist angle for (b).}
\label{SI_FIG1}
\end{center}
\end{figure*}

\begin{table}[htp]
\begin{center}
\begin{tabular}{lcccccccc}
(n,m) $\quad$ & $\theta$ / degree $\quad$ & (1) & $\quad$ & (2) & $\quad$ & (3) & $\quad$ & (4)\\
\hline\hline
(15,16) & 2.13  & 418.9 & & 415.6  & & 414.5  & & 417.0  \\
(19,20) & 1.70  & 339.8 & & 336.3  & & 329.4  & & 330.1  \\
(21,22) & 1.54  & 310.3 & & 308.1  & & 313.8  & & 312.0  \\
(23,24) & 1.41  & 283.3 & & 283.7  & & 291.2  & & 285.4  \\
(25,26) & 1.29  & 260.6 & & 267.7  & & 260.3  & & 260.8  \\
(26,27) & 1.25  & 249.1 & & 247.8  & & 253.3  & & 250.6  \\
(27,28) & 1.20  & 234.2 & & 232.6  & & 249.2  & & 249.2  \\
(28,29) & 1.16  & 232.4 & & 236.9  & & 237.4  & & 237.6  \\
(29,30) & 1.12  & 224.5 & & 224.0  & & 229.3  & & 225.8  \\
(31,32) & 1.05  & 201.8 & & 206.9  & & 210.7  & & 210.3  \\
\end{tabular}
\end{center}
\caption{On-site Hubbard parameters, in units of meV, calculated from a Coulomb potential with $\epsilon_{r} = 1$.}
\label{OHP}
\end{table}

\begin{table}[htp]
\begin{center}
\begin{tabular}{lcccccccc}
(n,m) $\quad$ & $\theta$ / degree $\quad$ & (1) & $\quad$ & (2) & $\quad$ & (3) & $\quad$ & (4)\\
\hline\hline 
(15,16) & 2.13  & 258.4 & & 255.4  & & 254.2  & & 256.6  \\
(19,20) & 1.70  & 188.1 & & 185.0  & & 179.1  & & 179.6  \\
(21,22) & 1.54  & 163.7 & & 161.5  & & 164.1  & & 159.2  \\
(23,24) & 1.41  & 142.0 & & 142.1  & & 148.6  & & 143.4  \\
(25,26) & 1.29  & 124.6 & & 129.9  & & 122.8  & & 123.4  \\
(26,27) & 1.25  & 116.1 & & 115.4  & & 123.6  & & 116.6  \\
(27,28) & 1.20  & 105.8 & & 105.1  & & 116.5  & & 116.5  \\
(28,29) & 1.16  & 104.8 & & 107.9 & & 108.3  & & 108.5  \\
(29,30) & 1.12  & 99.7 & & 99.0  & & 102.8  & & 100.2  \\
(31,32) & 1.05  & 85.7 & & 88.6  & & 90.9  & & 90.7  \\
\end{tabular}
\end{center}
\caption{On-site Hubbard parameters, in units of meV, calculated in the presence of a metallic gate with $\epsilon_{r} = 1$.}
\label{OHPS}
\end{table}

\begin{table}[t]
\begin{center}
\begin{tabular}{lrr}
Initial Guess $\quad$ & $\quad$ $m_{00}$ / meV/degree & $\quad$ $c_{00}$ / meV \\
\hline\hline 
(1)     &   198.8    &    -   \\
        &   198.8    &   -0.1    \\
\hline
(2)     &    198.8   &    -    \\
        &   192.0    &    9.7   \\
\hline
(3)     &    200.4   &    -    \\
        &   184.0    &    24   \\
\hline
(4)     &   199.8    &    -    \\
        &   187.1    &    18.6   \\
\end{tabular}
\end{center}
\caption{Gradients and intercepts from linear fits of the on-site Hubbard parameters calculated from a Coulomb potential with $\epsilon_{r} = 1$. Top line of each initial guess is fitted with a line forced through the origin; while the second line is fitted with a free intercept.}
\label{CIOHP}
\end{table}

\begin{table}[t]
\begin{center}
\begin{tabular}{lrr}
Initial Guess $\quad$ & $\quad$ $m^{g}_{00}$ / meV/degree & $\quad$ $c^{g}_{00}$ / meV \\
\hline\hline 
(1)     &   103.0    &    -   \\
        &   159.5    &    -82.1   \\
\hline
(2)     &   102.8    &    -    \\
        &   154.0    &    -74.5   \\
\hline
(3)     &    104.0   &    -    \\
        &   147.1    &   -62.8    \\
\hline
(4)     &   103.0    &    -    \\
        &   145.0    &   -68.4    \\
\end{tabular}
\end{center}
\caption{Gradients and intercepts from linear fits of the on-site Hubbard parameters calculated in the presence of a metallic gate with $\epsilon_{r} = 1$. Top line of each initial guess is fitted with a line forced through the origin; while the second line is fitted with a free intercept.}
\label{CHOHPS}
\end{table}

\clearpage

\section{Extended Hubbard Parameters}

All of the initial guesses for the extended Hubbard parameters were from Ref. \citenum{SMLWF}. All extended Hubbard parameters were calculated by displacing the first Wannier function along one of the lattice vectors of the system (either $\textbf{t}_{1}$ or $\textbf{t}_{2}$, but not a combination of both). There was always an integer, from 1 to 5, multiply the lattice vector. 

\begin{itemize}
    \item (1) - Interaction from the same AB position along $\textbf{t}_{1}$
    \item (2) - Interaction from the same BA position along $\textbf{t}_{1}$
    \item (3) - Interaction from the same AB position along $\textbf{t}_{2}$
    \item (4) - Interaction from the same BA position along $\textbf{t}_{2}$
    \item (5) - Interaction between AB and BA position along $\textbf{t}_{1}$
    \item (6) - Interaction between BA and AB position along $\textbf{t}_{1}$
    \item (7) - Interaction between AB and BA position along $\textbf{t}_{2}$
    \item (8) - Interaction between BA and AB position along $\textbf{t}_{2}$
\end{itemize}

By comparing (1) and (2), for example, the similarity between the two calculated Wannier functions can be determined. The extended Hubbard parameters calculated from these different locations should be identical~\cite{MLWO,SMLWF}, but because two Wannier functions were selectively localised with constrained centres, there are small differences between the values of the parameters. 

\clearpage

\subsection{Extended Hubbard Parameters - (19,20)}

\begin{table}[htp]
\begin{center}
\begin{tabular}{l ccccccc cccccccc}
$r$ / $\textrm{\AA}$ $\quad$ &  (1) & $\quad$ & (2) & $\quad$ & (3) & $\quad$ & (4) & $\quad$ & (5) & $\quad$ & (6) & $\quad$ & (7) & $\quad$ & (8)\\
\hline\hline 
48.0    & - && - && - && - && 272.6 & & -  & & 274.1  & & -  \\
83.1    & 200.3 & & 202.1  & &  205.3 & & 200.8 && - && - && - && - \\
127.0   & - && - && - && - && 126.8 & & 127.6  & & 127.6  & & 129.3  \\
166.2   & 92.7 & & 93.0  & &  94.6 & & 92.8  && - && - && - && - \\
209.1   & - && - && - && - && 71.6 & & 71.7  & &  71.8 & & 72.2  \\
249.3   & 59.3 & & 59.4  & &  59.7 & & 59.3  && - && - && - && - \\
291.8   & - && - && - && - && 50.2 & & 50.3  & & 50.3  & & 50.4  \\
332.4   & 44.0 & & 44.0  & &  44.1 & & 43.9  && - && - && - && - \\
374.7   & - && - && - && - && 38.8 & & 38.9  & &  38.9 & & 38.9  \\
415.5   & 35.0 & &  35.0 & &  35.0 & & 25.0  && - && - && - && - \\
457.6   & - && - && - && - && -  & & 31.7  & & -  & & 31.7  \\
\end{tabular}
\end{center}
\caption{On-site Hubbard parameters, in units of meV, calculated from a Coulomb potential with $\epsilon_{r} = 1$ for a twist angle of 1.70$^{\text{0}}$.}
\end{table}

\begin{table}[htp]
\begin{center}
\begin{tabular}{l ccccccc cccccccc}
$r$ / $\textrm{\AA}$ $\quad$ &  (1) & $\quad$ & (2) & $\quad$ & (3) & $\quad$ & (4) & $\quad$ & (5) & $\quad$ & (6) & $\quad$ & (7) & $\quad$ & (8)\\
\hline\hline 
48.0    & - && - && - && - && 131.0 & & -  & & 132.4  & & -  \\
83.1    & 73.7 & & 75.3  & & 78.2  & & 74.1  && - && - && - && - \\
127.0   & - && - && - && - && 22.5 & &  23.2 & &  23.2 & & 24.6  \\
166.2   & 6.9 & &  7.0 & &  8.0 & & 6.9  && - && - && - && - \\
209.1   & - && - && - && - && 1.7 & & 1.8  & &  1.8 & & 2.0 \\
249.3   & 4.7x10$^{-1}$ & & 4.7x10$^{-1}$  & &  5.5x10$^{-1}$ & & 4.7x10$^{-1}$ && - && - && - && -  \\
291.8   & - && - && - && - && 1.1x10$^{-1}$ & & 1.2x10$^{-1}$  & & 1.1x10$^{-1}$  & &  1.4x10$^{-1}$ \\
332.4   & 3.0x10$^{-2}$ & & 3.1x10$^{-2}$  & & 3.5x10$^{-2}$  & &  3.0x10$^{-2}$ && - && - && - && - \\
374.7   & - && - && - && - && 7.3x10$^{-3}$ & & 8.1x10$^{-3}$  & & 7.4x10$^{-3}$  & &  9.0x10$^{-3}$ \\
415.5   & 2.0x10$^{-3}$ & & 2.0x10$^{-3}$  & & 2.3x10$^{-3}$  & &  2.0x10$^{-3}$ && - && - && - && - \\
457.6   & - && - && - && - && -  & &  5.4x10$^{-4}$ & & -  & & 6.0x10$^{-4}$  \\
\end{tabular}
\end{center}
\caption{On-site Hubbard parameters, in units of meV, calculated with account of metallic gate with $\epsilon_{r} = 1$ for a twist angle of 1.70$^{\text{0}}$.}
\end{table}

\clearpage

\subsection{Extended Hubbard Parameters - (25,26)}

\begin{table}[htp]
\begin{center}
\begin{tabular}{l ccccccc cccccccc}
$r$ / $\textrm{\AA}$ $\quad$ &  (1) & $\quad$ & (2) & $\quad$ & (3) & $\quad$ & (4) & $\quad$ & (5) & $\quad$ & (6) & $\quad$ & (7) & $\quad$ & (8)\\
\hline\hline 
62.7    & - && - && - && - && 216.5 & &  - & & 213.7  & &  - \\
108.6   & 156.7 & & 157.0  & & 159.1  & & 159.1 && - && - && - && -  \\
166.0   & - && - && - && - && 96.3 & &  97.2 & & 97.5  & &  98.5 \\
217.3   & 71.1 & &  71.2 & & 72.0  & &  71.6 && - && - && - && - \\
273.4   & - && - && - && - && 54.6 & & 54.9  & & 54.9  & &  55.2 \\
326.0   & 45.4 & &  45.4 & & 45.6  & &  45.5 && - && - && - && -  \\
381.6   & - && - && - && - && 38.4 & & 38.5  & & 38.5  & &  38.6 \\
434.6   & 33.6 & &  33.6 & & 33.7  & &  33.6 && - && - && - && - \\
490.0   & - && - && - && - && 29.7 & & 29.7  & & 29.7  & & 29.8  \\
543.3   & 26.7 & &  26.7 & & 26.8  & &  26.8 && - && - && - && - \\
598.4   & - && - && - && - && - & & 24.2  & & - & & 24.3  \\
\end{tabular}
\end{center}
\caption{On-site Hubbard parameters, in units of meV, calculated from a Coulomb potential with $\epsilon_{r} = 1$ for a twist angle of 1.29$^{\text{0}}$.}
\end{table}

\begin{table}[htp]
\begin{center}
\begin{tabular}{l ccccccc cccccccc}
$r$ / $\textrm{\AA}$ $\quad$ &  (1) & $\quad$ & (2) & $\quad$ & (3) & $\quad$ & (4) & $\quad$ & (5) & $\quad$ & (6) & $\quad$ & (7) & $\quad$ & (8)\\
\hline\hline 
62.7    &  - && - && - && - && 90.8 & & -  & &  87.4 & & -  \\
108.6   & 48.2 & & 48.2  & & 50.2  & & 50.1  && - && - && - && -\\
166.0   &  - && - && - && - && 11.0 & & 12.2  & & 12.1  & &  13.0 \\
217.3   & 3.1 & & 3.1  & & 3.6  & &  3.3 && - && - && - && -\\
273.4   &  - && - && - && - && 4.9x10$^{-1}$ & & 6.6x10$^{-1}$  & & 6.0x10$^{-1}$  & & 7.2x10$^{-1}$  \\
326.0   & 1.2x10$^{-1}$ & & 1.2x10$^{-1}$  & & 1.5x10$^{-1}$  & & 1.3x10$^{-1}$  && - && - && - && -\\
381.6   &  - && - && - && - && 1.6x10$^{-2}$ & & 2.7x10$^{-2}$  & & 2.0x10$^{-2}$  & & 3.0x10$^{-2}$  \\
434.6   & 3.9x10$^{-3}$ & & 4.0x10$^{-3}$  & & 4.8x10$^{-3}$  & & 4.1x10$^{-3}$  && - && - && - && -\\
490.0   &  - && - && - && - && 4.8x10$^{-4}$ & & 8.9x10$^{-4}$  & & 6.0x10$^{-4}$  & &  9.6x10$^{-4}$ \\
543.3   & 1.2x10$^{-4}$ & & 1.2x10$^{-4}$  & & 1.5x10$^{-4}$  & &  1.2x10$^{-4}$ && - && - && - && -\\
598.4   & - && - && - && - &&  -  & & 2.8x10$^{-5}$  & &  -  & & 3.0x10$^{-5}$  \\
\end{tabular}
\end{center}
\caption{On-site Hubbard parameters, in units of meV, calculated with account of metallic gate with $\epsilon_{r} = 1$ for a twist angle of 1.29$^{\text{0}}$.}
\end{table}

\clearpage

\subsection{Extended Hubbard Parameters - (31,32)}

\begin{table}[htp]
\begin{center}
\begin{tabular}{l ccccccc cccccccc}
$r$ / $\textrm{\AA}$ $\quad$ &  (1) & $\quad$ & (2) & $\quad$ & (3) & $\quad$ & (4) & $\quad$ & (5) & $\quad$ & (6) & $\quad$ & (7) & $\quad$ & (8)\\
\hline\hline 
77.5    & - && - && - && - && 173.4 & &  - & & 173.1  & & -  \\
134.2   & 128.4 & & 128.3  & & 128.2  & & 128.3 & & - && - && - && - \\
205.0   & - && - && - && - && 78.1 & & 78.3  & & 78.0  & & 78.2  \\
268.4   & 58.5 & & 58.5  & & 58.4  & & 58.4 & & - && - && - && - \\
337.8   & - && - && - && - && 44.8 & &  44.9 & & 44.6  & & 44.9  \\
402.7   & 37.0 & & 37.0  & & 37.0  & & 37.0 & & - && - && - && - \\
471.4   & - && - && - && - && 31.2 & &  31.3 & & 31.2  & &  31.3 \\
536.9   & 27.3 & & 27.3  & & 27.3  & & 27.3 & & - && - && - && - \\
605.2   & - && - && - && - && 24.1 & & 24.1  & & 24.1  & & 24.1  \\
671.1   & 21.7 & & 21.7  & & 21.7  & & 21.7 & & - && - && - && - \\
739.2   & - && - && - && - && - & & 19.7  & & -  & &  19.7 \\
\end{tabular}
\end{center}
\caption{On-site Hubbard parameters, in units of meV, calculated from a Coulomb potential with $\epsilon_{r} = 1$ for a twist angle of 1.05$^{\text{0}}$.}
\end{table}

\begin{table}[htp]
\begin{center}
\begin{tabular}{l ccccccc cccccccc}
$r$ / $\textrm{\AA}$ $\quad$ &  (1) & $\quad$ & (2) & $\quad$ & (3) & $\quad$ & (4) & $\quad$ & (5) & $\quad$ & (6) & $\quad$ & (7) & $\quad$ & (8)\\
\hline\hline 
77.5    & - && - && - && - && 64.3 & & -  & & 64.0  & &  - \\
134.2   & 35.3 & & 35.3  & & 35.2  & &  25.4 & & - && - && - && - \\
205.0   & - && - && - && - && 7.2 & & 7.2  & & 7.1  & &  7.2 \\
268.4   & 2.3 & & 2.3  & & 2.3  & & 2.3  & & - && - && - && - \\
337.8   & - && - && - && - && 4.6x10$^{-1}$ & & 5.6x10$^{-1}$  & & 4.5x10$^{-1}$  & & 5.6x10$^{-1}$  \\
402.7   & 1.4x10$^{-1}$ & & 1.4x10$^{-1}$  & & 1.3x10$^{-1}$  & & 1.4x10$^{-1}$  & & - && - && - && - \\
471.4   & - && - && - && - && 1.4x10$^{-2}$ & & 3.9x10$^{-2}$  & & 1.3x10$^{-2}$  & & 4.0x10$^{-2}$  \\
536.9   & 4.2x10$^{-3}$ & & 4.3x10$^{-3}$  & & 4.2x10$^{-3}$  & & 4.2x10$^{-3}$  & & - && - && - && - \\
605.2   & - && - && - && - && 3.6x10$^{-4}$  & & 1.3x10$^{-3}$  & &  3.4x10$^{-4}$  & & 1.4x10$^{-3}$  \\
671.1   & 1.2x10$^{-4}$ & & 1.2x10$^{-4}$  & & 1.1x10$^{-4}$  & & 1.2x10$^{-4}$  & & - && - && - && - \\
739.2   & - && - && - && - && - & &  4.0x10$^{-5}$  & & -  & & 4.1x10$^{-5}$ \\
\end{tabular}
\end{center}
\caption{On-site Hubbard parameters, in units of meV, calculated with account of metallic gate with $\epsilon_{r} = 1$ for a twist angle of 1.05$^{\text{0}}$.}
\end{table}

\end{document}